# CHALLENGES AND OPTIMIZATION OF Mu2e PROTON TARGET DESIGN WITH RADIATIVE COOLING*

Z. Liu[†,1], G. Annala[1], M. Bloomer[2], A. Edmonds[3], M. Hedges[1], K. Lynch[1], A. Makovec[1], J. Miller[4], F. Pellemoine[1], J. Popp[3], J. Williams[1], K. Yonehara[1]
[1]Fermi National Accelerator Laboratory, Batavia, IL, USA
[2]Emory University, Atlanta, GA, USA
[3]York College / CUNY, Jamaica, NY, USA
[4]Boston University, Boston, MA, USA

*Abstract*

The Mu2e experiment at Fermilab will search for the charged lepton flavour violating process of coherent neutrinoless muon-to-electron conversion in the presence of an aluminum nucleus. The muons are produced by an 8 GeV proton beam from the Fermilab Booster striking a production target to create hadrons that decay to muons. The production target design space is strongly constrained by a required one-year operating lifetime and the need for radiative cooling in a vacuum. Uncertainties in the lifetime of the existing baseline design – a monolithic, segmented tungsten (WL10) target – are large, particularly due to unknown effects of radiation damage at the very high proton fluences expected in the experiment. We have begun evaluating a new design utilizing Inconel 718. Here, we present an engineering analysis of a prototype modular design. specifically thermal management, structural stability, fatigue lifetime, and fabrication changes. The results approve a promising new target design for the Mu2e experiment.

## INTRODUCTION

The Mu2e experiment at Fermilab, will search for the coherent neutrinoless conversion of muons to electrons in aluminium [1]. An 8 GeV proton beam from Fermilab Booster strikes the production target to produce particles, mostly pions, that decay to muons for the experiment. Further details on the Mu2e experiment design, motivation, and proposed run plan can be found in Ref. [1].

The production target design is fundamental to the success of Mu2e experiment as the source of the muon beam. The existing production target design, codenamed "Hayman", shown in the top-left of Fig. 1, is made from tungsten, a high Z material, to maximize production while the geometry is designed to minimize reabsorption. The Hayman target is 220 mm long with its core of 6 mm in diameter and four fins to increase surface for radiation cooling. The target support structure consists of a ring with six spokes that attach to end-rings affixed to the monolithic target. The whole assembly of the ring and the target shall be removed for change through a remote handling system (RHS), no more than once per year.

Radiation damage to the production target is a key factor in ensuring lifetime requirements. Radiation damage changes physical properties of the material, including mechanical strength and thermal conductivity. The degree of radiation damage is often quantified with displacements per atom (dpa) at lattice level. As dpa increases, the material's ability to manage heat and absorb shock decreases, accelerating degradation and failure. Unfortunately, recent application of tungsten in highly irradiated conditions has shown that it's possibly harmfully susceptible to radiation damage [2]. For a tungsten target subjected to the 8 GeV proton beam at Mu2e for one year of running, preliminary simulation studies, using FLUKA [3, 4] show dpa levels approximately an order of magnitude above operational limits found in the literature [5], suggesting alternative target material considerations are warranted.

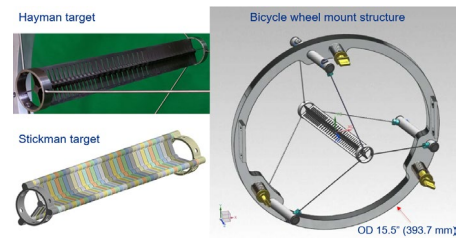

Figure 1: Mu2e production targets and mount structure.

Recently, a G4Beamline [6] simulation of muon production per proton on target (POT) shows that the muon production does not increase linearly with target density but does so with the nuclear interaction length [7]. This motivates investigation into previously unconsidered materials with available data on high levels of radiation damage. Nickel-based Inconel is one such option for an alternative material for the Mu2e target [8].

Inconel 600 was used in the Fermilab Antiproton Source [9], but there is no record of Post Irradiation Examinations (PIE) on this target. Recently a solution-annealed Inconel 718 proton beam window at the Spallation Neutron Source at Oak Ridge National Laboratory was reported with observation of an increase in ductility, rather than embrittlement, under 10 dpa of proton irradiation [10]. Preliminary physics analysis estimates a peak of 10 dpa in an Inconel 718 target in one-year of proton beam irradiation for Mu2e. In addition, Inconel 718 has high emissivity, up to 0.9 of hemispherical total emissivity [11], which suits radiation cooling as a requirement for the design. These motivate the design of an Inconel 718 target design, codenamed "Stickman" as shown in bottom left of Fig. 1.

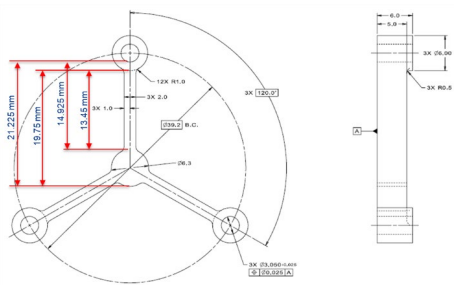

Figure 2: Stickman target plate (typical).

The Stickman target must fit within the confines of the RHS for safe removal and replacement of the target. To serve that purpose, the mount structure in Fig. 1 stays the same except minor changes such as Inconel 718 for the spokes. The Stickman has three fins rather than four fins, which is easier to deal with for assembly and mount. The Stickman target is of an assembly of thirty-five plates (Fig. 2), three bars, two end rings, and six spacers. It's expected to simplify the fabrication process and isolate failure of a single plate from affecting the rest when the target is fabricated by pieces. The overall length and core diameter of the Stickman target remain unchanged relative to the previous "Hayman" design.

## THERMAL MANAGEMENT

### Proton Beam for Mu2e

The macro-structure of the proton beam on the Mu2e proton target is illustrated in Fig. 3. Two Booster proton batches are each re-bunched into four spills of $1 \times 10^{12}$ protons with a kinetic energy of 8 GeV. For each spill, the proton target will see a train of proton pulses lasting 43.1 msec followed by a 5 msec reset period with no beam. This is repeated eight times during each cycle for a total beam on time of 380 msec. At the end of the eight spills, beam to Mu2e is off for 1020 msec. The maximum number of protons per spill is assumed to be $1.25 \times 10^{12}$, which is counted for thermal management as a conservative approach. The power per spill of the protons is about 1602 J within 43.1 msec.

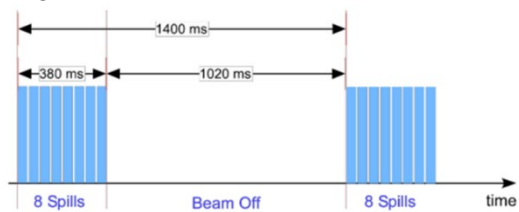

Figure 3: Time structure of the proton beam for Mu2e.

### Energy Deposition in Stickman Target

When the protons hit the target, part of the energy is deposited in the material through interaction. The energy deposition (Edep) is calculated with FLUKA, then imported into the commercial software ANSYS for finite element analysis (FEA); import data are mapped to the FEA meshing through profile fit.

### Thermal Analysis

The heat deposited in the target will dissipate through thermal radiation, which follows the Stefan-Boltzmann Law with the governing principle of Eq. (1).

$$P = \sigma \times \varepsilon \times A \times (T^4 - T_b^4) \quad (1)$$

where, $P$ is the energy deposition in the target from interaction with the protons, $\sigma$ the Stefan-Boltzmann constant, $\varepsilon$ the surface emissivity, $A$ the radiation surface, $T$ the temperature of the target, and $T_b$ the ambient temperature.

Calculation following Eq. (1) were performed for a rod of radius of 22.6 mm resembling the radiation surface of the target of thirty-five plates (Fig. 2), about 0.025 mm². With the power of about 400 W, the hemispherical total emissivity 0.85 at 600 °C (Fig. 4), and surrounding temperature 25 °C, the average target temperature is about 490 °C.

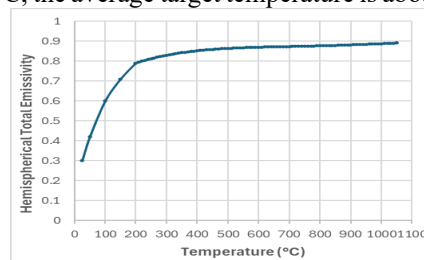

Figure 4: Hemispherical total emissivity of Inconel 718 oxidized at 1000 °C, with assumption of emissivity 0.3 at room temperature and curve fitting between 25 °C and data above 200 °C [11].

The temperature distribution within the target assembly is revealed through FEA. Temperature dependent material properties [12-15] are applied in the FEA, while Table 1 lists as references the corresponding properties at room temperature and at 648.9 °C / 1200 °F. After 14 s cycling from temperature distribution of steady-state FEA, as shown in Fig. 5, the peak temperature cycles with the proton beam on and off, indicating stability in terms of thermal balance. For a single cycle of heating in Fig. 6, peak temperature occurs in the target plate and low temperature 62 °C in the ring. On the right side is an enlarged sectional view of the 6th target plate showing distribution of temperature ranging from the maximum value 656.17 °C to 432.05 °C.

Table 1: Inconel 718 Properties (AMS 5596 sheet/plate, AMS 5662 and 5663 bars and rings) [12-15]

| Property | At 21 °C | At 649 °C |
|---|---|---|
| Density | 8.19 g/cc | 8.00 g/cc |
| Young's Modulus (E) | 200 GPa | 164 GPa |
| Torsional Modulus (G) | 77.2 GPa | 62.7 GPa |
| Poisson's Ratio (ν) | 0.294 | 0.283 |
| Tensile Strength | 1241 MPa | 965 MPa* |
| Yield Strength | 1034 MPa | 793 MPa* |
| Thermal Conductivity | 11.4 W/m-K | 21.2 W/m-K |
| CTE (α) | 12.3 μm/m-°C | 14.98 μm/m-°C |
| Specific heat $C_p$ | 428 J/kg-K | 595 J/kg-K |

*Future testing will give reliable strength data.

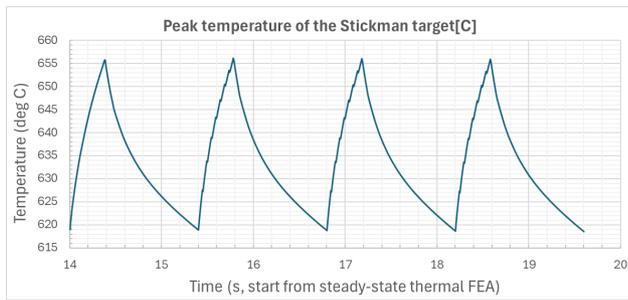

Figure 5: Peak temperature cycles with the heating and cooling. After 14 s cycling from temperature distribution of steady-state FEA, the first cycle at left is for heating evenly distributed within 380 ms, while the other three cycles for eight times of 43.1 ms beam-on and 5 ms beam-off and followed by 1.02 s beam-off.

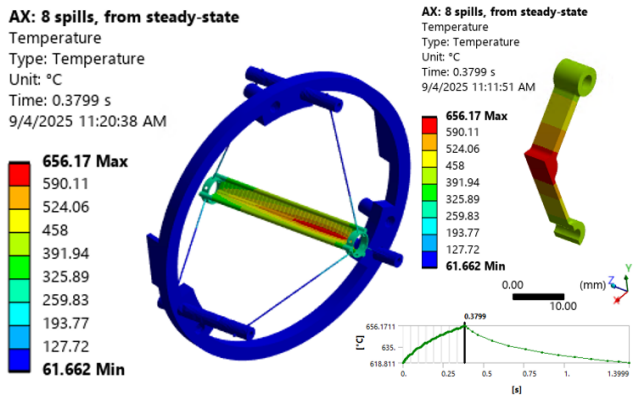

Figure 6: Temperature distribution of the target assembly. The enlarged section view of the plate with the maximum temperature is shown at right side.

## STRUCTURAL ANALYSIS

### Stress Analysis

For an isotropic elastic material of Young's modulus E, Poison's ratio ν, and CTE α, the thermal stress S within the solid [16] is described as Eq. (2).

$$S = \frac{E\alpha}{1-2\vartheta}(T - T_b) \qquad (2)$$

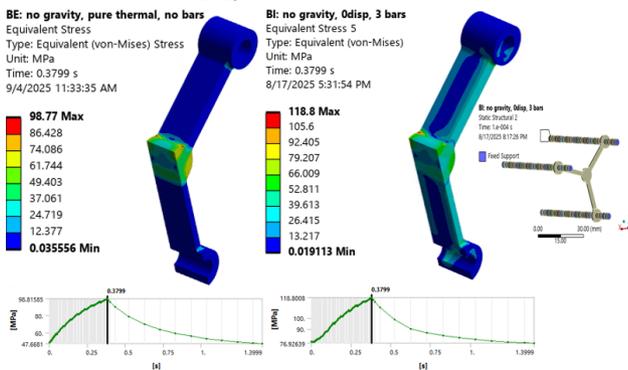

Figure 7: Stress of the target plate #6. Left: thermal stress without constraints. Right: thermal stress with fixed end surfaces of three rods and frictionless contacts.

When $T$ is greater than $T_b$ the constraints exert a compressive force on the material. A tensile force occurs when $T$ is less than $T_b$. Without constraints the thermal expansion or contraction leads to zero stresses. The zero stresses in the fins of the target plate (left image in Fig. 7) demonstrate no constraint is applied, while non-zero stresses in the core and corner are subject to geometrical constraints to the thermal expansion. In comparison the target plate has constraint from the support of three rods that are fixed at end faces (right image in Fig. 7), which results in non-zero stresses in the fins. The maximum value of von-Mises stress increases from 98.8 MPa to 118.8 MPa when the support is applied. At 650 °C the Inconel 718 yields at 793 MPa (Table 1), leading to a safety factor of the target plate of about 6.6 without counting radiation damage. When the first 8 spills of the proton beam hit the cold target, the peak temperature and maximum thermal stress rise to 92 °C and 75 MPa, lower than those in steady state.

### Fatigue, Rupture Life, and Creep Rate

The peak stress rises to 118.8 MPa and goes down to 76.9 MPa in one cycle of beams on and off, as shown on Fig. 7. With 1.4 s as the period of one cycle, the one-year operational lifetime for the target means $2.25 \times 10^7$ cycles per year. The cycle number and stress are much less than $10^8$ cycle under stress 582 MPa in Table 2 for the concern of fatigue. Similarly with the stress level, rupture life and creep rate are beyond concern as well if these properties won't be decreased by half due to radiation damage.

Table 2: Fatigue, Rupture Life and Creep Rate of Inconel 718 at 649 °C [12]

| Fatigue | | Rupture | | Creep | |
|---|---|---|---|---|---|
| Cycles | Strength | life | Stress | Rate | Stress |
| $10^8$ | 582 MPa | 8000 hr | 485 MPa | 0.000002 | 485 MPa |

## SUMMARY

Thermal and structural analyses of the optimized target design predict its performance. The maximum temperature is about 656 °C, above which the strength of Inconel 718 descends dramatically [12]. The maximum stress is at the level of 120 MPa, which leads to a safety factor of 6.6 with respect to its yield point. This stress level also predicts that the target may perform well within one-year operational lifetime, including more than one year of fatigue lifetime and rupture life with zero creep. Radiation damage to target material was considered for Mu2e production target design. The observation of an increase in ductility, rather than embrittlement, of a solution-annealed Inconel 718 proton beam window under 10 dpa of proton irradiation gives some confidence that the lifetime of the Inconel 718 target will not decrease drastically due to radiation damage. Radiation damage effects for material properties of Inconel 718 are ongoing. It's known that annealing and aging play important roles in material properties, which have been considered in design and fabrication. This prediction shows that these optimizations lead to a promising new target design for Mu2e experiment.